\documentclass[twocolumn,showpacs,preprintnumbers,amsmath,amssymb]{revtex4}

\usepackage{graphicx}
\usepackage{dcolumn}
\usepackage{bm}

\begin{document}

\title{Emergence of continual directed flow in Hamiltonian systems}

\author{D. Hennig}
\author{A.D. Burbanks}
\author{C. Mulhern}
\author{A.H. Osbaldestin}
\affiliation
{Department of Mathematics, University of Portsmouth, Portsmouth, PO1 3HF, UK}


\date{\today}

\begin{abstract}

\noindent We propose a minimal model for the emergence of a directed
flow in autonomous Hamiltonian systems. It is shown that internal
breaking of the spatio-temporal symmetries, via localised initial
conditions, that are unbiased with respect to the transporting degree
of freedom, and transient chaos conspire to form the physical
mechanism for the occurrence of a current.  Most importantly, after
passage through the transient chaos, trajectories perform solely
regular transporting motion so that the resulting current is of
continual ballistic nature. This has to be distinguished from the
features of transport reported previously for driven Hamiltonian
systems with mixed phase space where transport is determined by
intermittent behaviour exhibiting power-law decay statistics of the
duration of regular ballistic periods.
\end{abstract}

\maketitle

\noindent Nonlinear transport processes of particles evolving in a
spatially periodic potential have attracted considerable
interest~\cite{Risken} (for a recent review see \cite{review}). In
most of the studies, the emergence of particle current is triggered by
an external time-dependent field with zero mean (which can be of
stochastic nature) or is provided by a deterministic periodic force.
Recently, the Hamiltonian case has become the subject of intensive
studies due to its relevance to the motion of cold atoms in an optical
potential ~\cite{optics}.

The necessary conditions for rectification of the current, based on
symmetry investigations of the external field and the underlying
static potential, have been presented in~\cite{Flach}
and~\cite{Denisov}. To be precise, all symmetries that, to each
trajectory, generate a counterpart moving in the opposite direction,
need to be broken.  This is achievable by imposing a time-dependent
external force that is periodic but not symmetric under time
reversal~\cite{Flach}--\cite{Schanz1}.  Furthermore, the phase space
has to possess a mixing property, with coexisting regular and chaotic
dynamics~\cite{Schanz}.  In extended chaotic systems a nonzero current
can be obtained as the time-averaged velocity of an ensemble of
trajectories in the chaotic component of phase space, and the chaotic
transport proceeds ballistically and directedly~\cite{Schanz},
\cite{Schanz1}.

Extensions to studies of autonomous Hamiltonian systems of
one-dimensional billiard chains have followed~\cite{Acevedo},
\cite{Schanz2}. The necessity of creating chaos requires at least two
degrees of freedom. As an example of such a system, a classical
magnetic billiard for particles carrying an electric charge has been
studied in~\cite{Acevedo}.  In order to break the time-reversal
invariance, an external static magnetic field, penetrating the plane
of motion perpendicularly, has been applied. In addition, achieving
directed transport demands breaking the remaining spatial symmetry,
which can be achieved, e.g., by properly placed asymmetric obstacles
inside the billiard~\cite{Acevedo}, \cite{Schanz2}.  Uni-directional
motion in a serpent billiard chain has been reported in~\cite{Horvat}.

The aim of the current work is to demonstrate that chaotic directed
transport as achieved in systems with a mixed phase space is not the
only option to obtain directed transport in Hamiltonian systems. 
In driven Hamiltonian systems the directed transport necessitates a
mixed phase space for which chaotic trajectories can stick to the
boundaries of regular regions inducing long periods of nearly regular
motion so that finite asymptotic currents can be observed
\cite{Flach}--\cite{Schanz1}. On the other hand, sustained transport
seems to be impeded by the intermittent behaviour exhibiting
power-law decay statistics of the duration of periods of regular
motion.  Therefore, for the sake of stability and reliability of a
transporting regime it is desirable to find ways to accomplish
continual regular transport. As we show, when dealing with
nonintegrable systems, it is advantageous when chaos is only of
transient nature and serves to guide trajectories onto {\it regular
  transporting} motion.  Furthermore, we also show that, in contrast
to the studies quoted above, a directed flow can arise in autonomous
Hamiltonian systems even without the application of a time
reversibility symmetry breaking external field: specifically, we show
that while the system as a whole is time-reversible,
physically-relevant sets of localised initial conditions lead to
current despite being \emph{unbiased} with respect to the transporting
degree of freedom.

We consider a Hamiltonian system with $n$ degrees of freedom of the
form
\begin{equation}
H(p,q)=\frac{1}{2}p^2+U(q)\label{eq:hamiltonian}
\end{equation}
with $(p,q)\in \mathbb{R}^{2n}$ and $U(q)$ as a potential function.
It is assumed that the system possesses an open component, by which we
mean that constant energy surfaces may be unbounded in the
coordinate(s), allowing for transient chaos.

In the following, we discuss the spatio-temporal symmetry properties
of the Hamiltonian equations $\dot{p}=-\partial H/\partial q$ and
$\dot{q}=\partial H/\partial p$.  Firstly, the system of equations
exhibits time reversible invariance. Solutions are of the form
$X(t)=(p(t),q(t))$.  Applying the time-reversal operator yields
$\widehat{\tau}(p(t),q(t))=(-p(-t),q(-t))$ and, hence, if $X$ is a
solution, then so is $\widehat{\tau} X$. As for the implication of
time-reversibility symmetry with regard to the net flow, let a
solution, starting from some initial condition ${X}(0)$, be evolved in
time up to a finite observation time $T$ at which the {\it forward}
trajectory arrives at the point ${X}(T)$ on the constant energy
surface. Subsequently, under application of the time-reversal
operation, the signs of the momenta at this point, ${X}(T)$, are
reversed, and letting the solution evolve once again, with
$\widehat{\tau}X(T)$ as the initial condition, the corresponding {\it
  backward} trajectory traces back the path of the forward trajectory
in coordinate space. For a microcanonical ensemble, the initial
conditions ${X}(0)$ and $\widehat{\tau}{X}(T)$ are equally selected
points from the constant energy surface.  Thus, for systems with
time-reversibility symmetry and uniformly distributed initial
conditions populating the whole energy surface there is no preferred
direction of the flow preventing the emergence of a current.

However, energy surfaces that are unbounded (along the coordinates of
the open component) may not, in practise, be completely populated with
a finite set of initial conditions.  In this sense, any finite set of
initial conditions can be regarded as being {\it localised} in the
coordinates of the spatially-open system. Localised initial
coordinates are frequently used in applications such as for the
problem of a particle flow emerging when the particles are initially
trapped in a well of a spatially infinitely extended, multiple-well
potential (see below).  Therefore, for practical purposes, it is
supposed that the initial conditions for coordinates are localised in
the domain $-q_{j,l}\le q_{j}(0) \le q_{j,r}$ and $1\le j \le n$.  We
define a trajectory of the spatially-open system as {\it transporting}
(with respect to a specified observation time, $T$) if: (i) at least
one of its coordinates, $q_{j}(t)$, escapes from the domain of the
localised initial conditions at some instant of time $0 < t_{escape}
\ll T$ and, (ii) performs subsequently directed net motion, that is
$\langle p_j(t)\rangle \ne 0$ for $T\ge t \ge t_{escape}$, where
$\langle \,\cdot\,\rangle$ denotes the time average. This leads the
trajectory away from the domain of localised initial conditions so
that at the end of the observation time, at least one of the terminal
coordinates of a transporting forward trajectory lies outside the
domain of the localised initial values; ${q}_j(T)<-q_{j,l}$ or
${q}_j(T)>q_{j,r}$.  Consequently, the initial condition of the
corresponding backward trajectory (which would compensate the
contribution of the forward trajectory to the net current) is not
contained in the set of localised initial conditions. It needs to be
stressed that this alone does not imply the emergence of a current in
the system for such sets of initial conditions.

Furthermore, although the equations of motion are time reversal
symmetric, not all of their solutions necessarily obey this
symmetry. Regarding the selection of an initial condition, $X(0)$, it
holds that unless a trajectory is self-reversed, i.e., for the
selected initial condition and its time-reversed counterpart,
$\widehat{\tau} X(0)$, the corresponding motion coincides in phase
space, i.e., $X=\widehat{\tau}X$, time reversibility is broken. In
fact, selecting an initial condition, $X(0)$, may break the
time-reversibility symmetry, which is the case when the trajectory $X$
and its time reversal counterpart $\widehat{\tau}X$ are
distinct. Examples of time-reversibility breaking are provided by the
(unbounded) rotating trajectories of a pendulum whereas the (bounded)
librating trajectories are self-reversed and thus time reversible
symmetric.  For potential systems of the form of
Eq.\,(\ref{eq:hamiltonian}), time-reversibility is manifested in
coordinate space in the spatial symmetry features induced by
reflections on the time-reversal symmetry hypersurfaces on which the
time evolution of the trajectories satisfies $p(-t)=-p(t)$ and
$q(-t)=q(t)$.  They are represented by $(n-1)$-dimensional manifolds
in the coordinate space and are determined by the condition
$\dot{p}_k=0$.  This gives
\begin{eqnarray}
S_{k}\,:&&\hspace{0.2cm} \frac{\partial U}{\partial q_k}
=F_k(q)
=0\,,\qquad 1 \le k \le n\,.\label{eq:sk}
\end{eqnarray}
(For the sake of illustration, we consider the case where each
condition in (\ref{eq:sk}) has a single solution and the resulting
symmetry manifolds intersect in a single point representing an
equilibrium of the system.)  Let us consider reflections of a
trajectory, projected onto coordinate space, on the symmetry manifolds
$S_k$ induced by the corresponding operators $\widehat{R}_{k}$.
Self-reversed trajectories are left invariant upon reflections on the
symmetry manifolds, $\prod_{k=1}^n \,\widehat{R}_{k}(q)=q$, since they
are mapped pointwise onto themselves on equipotentials
$U(q)=U(\widehat{R}_k(q))$ such that the sign on the r.h.s. in the
equations of motion for $\dot{p}_k$ is reversed, $F_k(q) \rightarrow
-F_k(\widehat{R}_{k}(q))$, $\,1\le k \le n$.  Observe that upon
reflecting on all symmetry manifolds the relation
$U(q)=U(\prod_{k=1}^m \,\widehat{R}_{k}(q))$, $1\le m \le n$ is left
invariant under permutations of the reflection operators.  In fact,
time-reversing symmetry is in coordinate space tantamount to
invariance with respect to reflections on the symmetry manifolds. In
more detail, any self-reversed trajectory, projected onto coordinate
space, repeatedly crosses every symmetry manifold $S_k$ each time with
an opposite sign of the corresponding force $-F_k(q) <
\infty$. Moreover, successive crossings of a single symmetry manifold,
$S_k$, occur in alternating directions. Thus there must be turning
points for the trajectory implying bounded motion and no directed flow
can arise.  Notice that no assumptions with regard to the spatial
symmetries of the trajectory are needed.  In contrast, as transporting
(unbounded) trajectories are not invariant with respect to reflections
on the symmetry manifolds, preservation of time-reversing symmetry is
not possible for such a single trajectory. A transporting trajectory
may escape without having crossed a symmetry manifold at all. However,
if it does cross then after all such crossings of a symmetry manifold,
the escaping trajectory promotes directed transport. Nevertheless,
reflections on the symmetry manifolds, mapping a transporting
trajectory onto another transporting one, can induce spatial
symmetries such that these two trajectories mutually compensate each
others contribution to the net flow. Let the point $q_O$ in coordinate
space be an initial condition associated with a transporting
trajectory.  Reflecting in coordinate space on the symmetry manifolds
$S_{k}$ transforms an original point, $q_O$, into its image point,
$q_I$, according to $\widehat{R}_{k}q_O=q_{I,k}$.  While the value of
the potential energy is maintained, $U(q_O)=U(q_{I,k})$, the sign on
the r.h.s. in the equations of motion for $\dot{p}_k$ is reversed;
$F_k(\widehat{R}_{k}q_O) \rightarrow -F_k(q_{I,k})$.  However, the
magnitude of the gradients, $F_k=\partial U/\partial q_k$, is not
necessarily maintained.  Reflection on all of the symmetry manifolds
yields $\prod_{k=1}^n \,\widehat{R}_{k}{q}_O=q_I$, reversing the sign
in all of the r.h.s. of the equations of motion for the evolution of
the momenta $F_k(q_I)\rightarrow -F_k(q_O)$,\,$1\le k \le n$.  With
the time evolution of a coordinate expressed as $q(t)=q(0)+\int_0^t
dt^{\prime}\{p(0)+\int_0^{t^{\prime}} dt^{\prime \prime} [-F(q(t
  ^{\prime \prime}))]\}$, we conclude that, for the pair of
trajectories emanating from $q_O$ and $q_I$, symmetry (zero net flow)
results if $(p_I,q_I)=(-p_O,-q_O)$ so that $F_k(q_I)=
-F_k(q_O)$,\,$1\le k \le n$. This is the case when the potential is
even in the coordinates, that is $U(q)=U(-q)$.  Then there exist pairs
of current-annihilating {\it counterpropagating} trajectories,
${X}(t)$, starting from ${X}(0)$, and $-{X}(t)$, starting from
$-{X}(0)$, respectively. In other words, {\it reversion symmetry}
under reflections on the symmetry manifolds is needed for zero net
flow which, together with invariance with respect to changes of the
sign of the momenta, amounts to parity-symmetry of the system
$H(p,q)=H(-p,-q)$.  Conversely, violation of reversion symmetry with
respect to at least one of the coordinates $q_k$ establishes a
prerequisite for the occurrence of directed flow.

For an illustration, we consider the conservative and deterministic
dynamics of a particle whose coordinate $q$ evolves in a
one-dimensional periodic, spatially-symmetric washboard potential of
unit period which is given by
\begin{equation}
U(q)=U(q+1)=\frac{1-\cos(2\pi q)}{2\pi}\,.\label{eq:U0}
\end{equation}
The particle is assumed to interact with a local oscillator of
amplitude $Q$ evolving in a harmonic potential being subjected to a
tilt force of strength $F$
\begin{equation}
V(Q)=\frac{1}{2}\omega^2Q^2+F\,Q+\frac{1}{2}\left(\frac{F}{\omega}\right)^2\,.\label{eq:V}
\end{equation}
Coupling between the particle and the harmonic oscillator arises from
the interaction potential
\begin{equation}
W(q,Q)=D\left[1-\frac{1}{\cosh(q-Q)}\right]\,,\label{eq:W}
\end{equation}
that depends on the relative distance $|q-Q|$. The parameter $D$
regulates the strength of the coupling between the particle and the
harmonic oscillator degree of freedom. Most importantly, the
interaction is of local character, having negligible influence on the
motion outside the so called \emph{interaction region}, in which when the
relative distance $|q-Q|$ is sufficiently small so that a
significantly large gradient of the interaction potential results.
The system of coupled equations of motion is given by
\begin{eqnarray}
\ddot{q}&=&-\sin(2 \pi q)-D\frac{\tanh(q-Q)}{\cosh(q-Q)}\label{eq:system1}\\
\ddot{Q}&=&-Q-F+D\frac{\tanh(q-Q)}{\cosh(q-Q)}\label{eq:system2}\,.
\end{eqnarray}
For $D=0$, the system decouples into two integrable subsystems and the
dynamics is characterized by individual regular motions of the
particle in the washboard potential, and bounded oscillations of the
harmonic degree of freedom, respectively. For $D\ne 0$, the subsystems
interact, exchanging energy.  While the $Q$-oscillator performs solely
bounded motion, there is the possibility that, for an escaping
particle, the corresponding coordinate, $|q|$ (representing the
spatially-open component), attains large values and thus the related
interaction forces, $\partial W/\partial q$ and $\partial W/ \partial
Q$, vanish asymptotically, allowing transient
chaos~\cite{Bleher}-\cite{Zaslavsky}.  That is, for large distance
$|q-Q|\gg 1$, the interaction vanishes asymptotically, with the result
that the two degrees of freedom effectively decouple, rendering the
dynamics regular. (Although this decoupling is a feature of the
specific example taken here, it is not essential for current emergence
in general.) 

We focus interest on the situation when the particle is initially at
rest at the bottom of one well of the washboard potential with initial
condition $p(0)={q}(0)=0$ (representing localised initial conditions
in the spatially-open component, situated in the centre of the
interaction region).  The total energy is initially deposited in the
harmonic degree of freedom. The particle can then escape from the well
only if it gains sufficient energy from its interaction with the
excited harmonic degree of freedom. The total energy is fixed at
$E=1.625$, which exceeds several times the barrier height,
$E_b=1/\pi\simeq 0.318$, of the washboard potential.
\begin{figure}
\includegraphics[height=7cm, width=8cm]{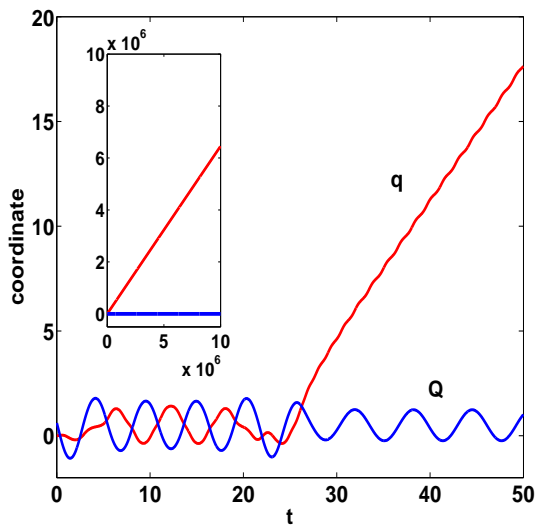}
\caption{ Typical time evolutions of the coordinates as indicated in
  the plot for $D=0.94$, $F=-0.5$, and $\omega=1$. Note the time scale
  on the inset figure; $0\le t\le 10^7$.} \label{fig:single}
\end{figure}
A typical trajectory is shown in Fig.~\ref{fig:single}.  Strikingly,
the particle gains sufficient energy from the harmonic degree of
freedom that it manages to escape from the potential well and
subsequently, avoids becoming trapped in wells of the washboard
potential indefinitely. Thus, the particle moves directedly to the
right (continually increasing the coordinate, $q$) while the
amplitude, $Q$, of the harmonic degree of freedom, performs small
oscillations around the bottom of the potential $V(Q)$. In other
words, after a chaotic transient the particle departs from the
interaction region and the dynamics of the two degrees of freedom
settles on individual regular motion.  (That the particle does,
indeed, settle on regular motion, and that the observed behaviour is
not merely a long-lived transient of the system, is proved in what
follows.)

We emphasise that the coordinate of the washboard particle tends to
infinity and the interaction term in the equations of motion vanishes
exponentially, excluding the return of the trajectory to the region of
the initial conditions, as we prove below.  (Remarkably, already at
$t=50$ the magnitude of the interaction term $D\tanh(q-Q)/\cosh(q-Q)$
has fallen below $10^{-7}$.)  In other words, the Poincar\'{e}
recurrence time is infinity.

\begin{figure}
\includegraphics[height=7cm, width=8cm]{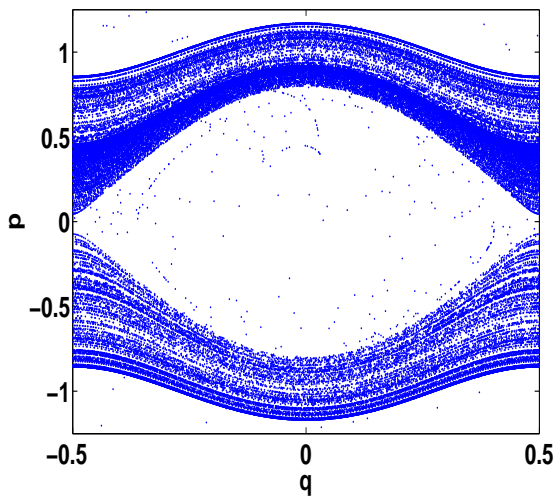}
\caption{Poincar\'e surfaces of section represented in the
  $(p,q)$-plane for $D=0.94$, $F=-0.5$, and
  $\omega=1$.} \label{fig:PSS}
\end{figure}

In order to illustrate the dynamics of an ensemble, evolving in the
four-dimensional phase space on the three-dimensional energy
hypersurface, we invoke a Poincar\'{e} surface of section taken as
$\Sigma=\left\{\,q, p\,|Q=0, P>0\,\right\}$.  Fig.~\ref{fig:PSS} shows
the Poincar\'e surface of section for $D=0.94$ and $F=-0.5$.
For the computation an ensemble of $10^5$ initial momenta and
coordinates $(P(0),Q(0))$ of the harmonic degree of freedom, is
taken. These initial values are uniformly distributed on the level
curve $E=0.5P^2+V(Q)+ W(0,Q)$ in the $(P,Q)$-plane.  Note that the
initial conditions are \emph{unbiased}, obeying the symmetry
$P\leftrightarrow -P$, and possess spatial symmetry with regard to the
washboard potential.  The simulation time interval is $T=10^5$ being
equivalent to almost $4\times10^4$ times the period of harmonic
oscillations near the bottom of a well of the washboard potential as
well as of the potential of the harmonic oscillator.  Likewise, the
simulation time exceeds by far the time it takes for the particle to
escape from the potential well.  The Poincar\'e surface of section,
shown in Fig.~\ref{fig:PSS}, is characterized by a few scattered
points related with the dynamics of chaotic transients (followed by
the trajectories during the particle's escape process) and densely
covered curves associated with the regular, rotational motion to which
the dynamics eventually adjusts. Crucially, there are in fact far more
trajectories evolving in the range of positive momenta than in the
negative range. Hence, net motion to the right arises.

It is illustrative to consider the symmetry properties underlying the
equations of motion, Eqs.\,(\ref{eq:system1}) and
(\ref{eq:system2}). Their structure is determined by the potential
functions given in Eqs.\,(\ref{eq:U0})--(\ref{eq:W}), which together
give rise to an effective potential
$U_\mathrm{eff}(q,Q)=U(q)+V(Q)+W(q,Q)$.  Regarding only the washboard
potential, it is periodic in $q$ with period $1$ and invariant under
reflection in $q$, i.e., $U(q)=U(-q)$.  This amounts to symmetry with
respect to every $q_n=n/2$ with integer $n$.  The interaction
potential is invariant with respect to changes of the sign of its
argument, viz., $(q-Q) \leftrightarrow -(q-Q)$. In contrast, the
tilted harmonic oscillator potential $V(Q)$ is without reflection
symmetry and, consequently, the system of Eqs.\,(\ref{eq:system1}) and
(\ref{eq:system2}) is not invariant under inversions $Q
\leftrightarrow -Q$.
\begin{figure}
\includegraphics[height=7cm, width=8cm]{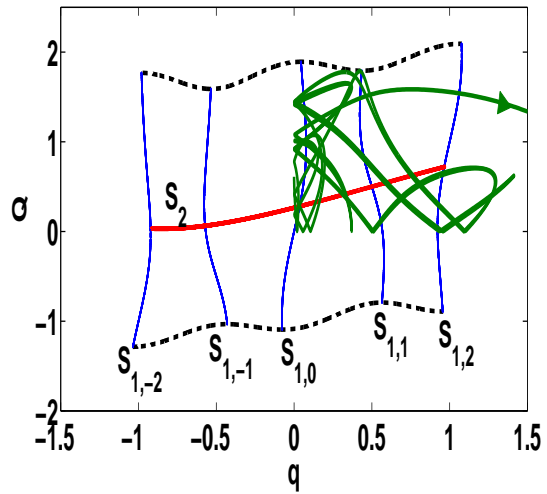}
\caption{Typical time evolutions of the coordinates as indicated in
  the plot for $D=0.94$, $F=-0.5$, and $\omega=1$.  The total energy
  is $E=1.625$ and the boundaries of the energetically allowed region
  are indicated by the horizontal dashed lines.  Superimposed is the
  trajectory shown in Fig.~\ref{fig:single}.} \label{fig:diagram}
\end{figure}
The symmetry lines are given by
\begin{eqnarray}
 S_1\,:&&\qquad \sin(2\pi q)+D\frac{\tanh(q-Q)}{\cosh(q-Q)}=0\label{eq:s1}\\
S_2\,:&&\qquad Q+F-D\frac{\tanh(q-Q)}{\cosh(q-Q)}=0\label{eq:s2}\,.
\end{eqnarray}
The symmetry line $S_1$ exhibits the following symmetry:
\begin{equation}
 Q\,\rightarrow -Q\,,\qquad \frac{n}{2}+q\,\rightarrow -\frac{n}{2}-q\,:\qquad S_{1,n}
 \rightarrow -S_{1,-n}
\label{eq:symmetry1}
\end{equation}
with integer $n$ labelling the several branches of the symmetry line
as $S_{1,n}$.  A single branch, $S_2$, arises from the second equation
with no apparent symmetry as shown in coordinate space in
Fig.~\ref{fig:diagram}.  The branches of the symmetry lines
$S_{1,n}$ related to the starting potential well and its neighbour to
either side are displayed in the figure, i.e., those with
$n=-2,-1,0,1,2$.  Crucially, reflections in the $(q,Q)$-coordinate
plane on the symmetry line $S_2$ do not respect inversion symmetry,
and thus, no pairs of counterpropagating trajectories are produced by
the localised, but otherwise unbiased, set of initial
conditions. Symmetry breaking of the trajectory displayed in
Fig.~\ref{fig:single} becomes apparent by superimposing the latter on
the diagram in coordinate space in Fig.~\ref{fig:diagram}. In
particular after an ultimate crossing of the symmetry line $S_{1,2}$
the trajectory proceeds exclusively to the right as indicated by the
arrow.  This spatial symmetry breaking is a prerequisite for the
occurrence of directed net motion.  In conclusion, even though the
initial conditions are unbiased for the ensuing dynamics, the
time-reversibility symmetry is broken.

We now prove that there exist trajectories which, once they have
entered a certain region in phase space, provide continual transport
of washboard particles.  The Hamiltonian associated with the system in
Eqs.\,({\ref{eq:system1}}),({\ref{eq:system2}) reads as
\begin{eqnarray}
 H&=&\frac{1}{2}p^2+U(q)+\frac{1}{2}P^2+V(Q)+W(q,Q)\nonumber\\
&\equiv&H_1(p,q)+H_2(P,Q)+W(q,Q)\,,\label{eq:hamiltonian1}
\end{eqnarray}
with $U,V,W$ given in (\ref{eq:U0}),(\ref{eq:V}),(\ref{eq:W}).

{\bf Theorem:} Consider the Cauchy problem for the Hamiltonian
(\ref{eq:hamiltonian1}) with initial data $(p(0),q(0),P(0),Q(0))$.

Let $E$ denote the total energy of the system and let $\widehat{Q}$
denote the positive root of the equation $V(\widehat{Q})=E$, i.e., the
maximal coordinate that the oscillator could attain if all of the
energy were in the oscillator degree of freedom.

Assume that the energies of the washboard particle and of the harmonic
oscillator satisfy the constraints
\begin{equation}
 E_1(0)\ge E_a+D\frac{1}{\cosh(q(0)-\widehat{Q})}\,,\label{eq:E10}
\end{equation}
and 
\begin{equation}
 E_2(0)<E-D-E_a\,,\label{eq:E20}
\end{equation}
respectively, where 
\begin{equation}
E_a>\frac{1}{\pi}\,,
\label{eq:Ea}
\end{equation}
the latter being the energy barrier height of the washboard degree of
freedom.

If, in addition, the intial values of the coordinates of the washboard
and oscillator satisfy the constraint
\begin{equation} 
 q(0)-Q(0)\ge q(0)-\widehat{Q}>\sinh^{-1}(1)\,,\label{eq:qQ}
\end{equation}
then it holds that, for some $\widehat{p}$,
\begin{equation}
 p(t)\ge\widehat{p}>0\,,
\end{equation}
for all times $t>0$.

{\bf Proof:} The proof utilises that existence of solutions with $p>0$
is guaranteed by the asymptotic behaviour of the washboard coordinate
as $q\rightarrow \infty$.  The momentum $p$ is represented as follows:
\begin{equation}
\frac{1}{2}p^2=E-U(q)-E_2-W(q,Q)\,.
\end{equation}
With the help of (\ref{eq:E20}) we obtain the lower bound for the
initial value
\begin{eqnarray} 
\frac{1}{2} p^2(0)
&=&
E_a-\max[U(q(0))]+D\frac{1}{\cosh(q(0)-Q(0))}\nonumber\\
&\ge&
E_a-\frac{1}{\pi}+D\frac{1}{\cosh(q(0)-Q(0))}\,.
\end{eqnarray}
From the lower inequality (\ref{eq:Ea}) follows that the momentum
$p(0)$ is positive, regardless of the value of the positive term
$D/\cosh(q(0)-Q(0))$.

The change in energy of the two subsystems under the dynamics is determined by 
\begin{eqnarray}
 \frac{dE_1}{dt}&=&\left\{E_1, H\right\}
 =-D\frac{\tanh(q-Q)}{\cosh(q-Q)} p\,,\\
 \frac{dE_2}{dt}&=&\left\{ E_2, H \right\}
 =D\frac{\tanh(q-Q)}{\cosh(q-Q)} P\,.
\end{eqnarray}
Suppose (in order to reach a contradiction) that the trajectory
reaches a point at which $p=0$, i.e., that the trajectory reaches a
stationary point in the washboard coordinate, $q$, after which it
might turn back. Let $q_1>q_0$ denote the first point at which that
happens.  Using that $p=dq/dt$ and $P=dQ/dt$ one obtains
\begin{eqnarray}
E_1(1)-E_1(0)
=-D\int_{q_0}^{q_1}\frac{\tanh(q-Q)}{\cosh(q-Q)}dq\,,\label{eq:int1}\\
E_2(1)-E_2(0)
=D\int_{Q_0}^{Q_1}\frac{\tanh(q-Q)}{\cosh(q-Q)}dQ\,.\label{eq:int2}
\end{eqnarray}
The maxima of the integrand ${\tanh(q-Q)}/{\cosh(q-Q)}$ are located at
$\lvert q-Q\rvert= \sinh^{-1}(1)$ and for 
\begin{equation}
\lvert q-Q \rvert >\sinh^{-1}(1)\label{eq:boundqQ}
\end{equation}
the integrand remains non-negative. Hence, provided the coordinates
stay in the range given by (\ref{eq:qQ}) the washboard degree of
freedom loses energy while the harmonic oscillator gains energy.  We
derive an upper bound on the magnitude of the energy loss of the
washboard particle and demonstrate that, provided the above
assumptions on the initial data hold, this loss cannot in fact be
sufficient to halt or turn back the particle.  To this end, the
integration in (\ref{eq:int1}) is performed yielding
\begin{eqnarray}
E_1(1)-E_1(0)
&=&
-D  \int_{q_0}^{q_1}\frac{\tanh(q-Q)}{\cosh(q-Q)}dq \label{eq:estimateexch}\\
&\ge&
-D \int_{q_0}^{q_1}\frac{\tanh(q-\widehat{Q})}{\cosh(q-\widehat{Q})}dq\nonumber\\
&=&
-D\left[-\frac{1}{\cosh(q_1-\widehat{Q})}+\frac{1}{\cosh(q_0-\widehat{Q})}\right]\nonumber\\
&\ge&
-D\left[\frac{1}{\cosh(q_0-\widehat{Q})}\right]\,.
\end{eqnarray}
Suppose that initial data for the coordinate $q(0)$ satisfies the
inequality (\ref{eq:qQ}) and, in addition, the washboard degree of
freedom satisfies our energy constraints (\ref{eq:E10}) and
(\ref{eq:Ea}), then
\begin{eqnarray}
E_1(1)
&\ge&
E_1(0)-D\left[\frac{1}{\cosh(q_0-\widehat{Q})}\right]\\
&\ge&
E_a>\frac{1}{\pi}\,.
\end{eqnarray}
Thus, the energy of the particle in the washboard degree of freedom
remains above the separatrix level, contradicting the supposition that
the particle halts (attaining $p=0$ at some position $q_1$), and we
therefore have that $p(t)>0$ for all times $t\ge 0$. Indeed, since the
energy $E_1$ of the washboard particle is bounded away from the barrier
level, $1/\pi$, it
follows that $p(t)\ge\widehat{p}>0$ for some $\widehat{p}$, which
implies that $q\to\infty$ as $t\to\infty$. \hfill
$\square$

That regions of phase space satisfying the assumptions of the theorem
do, indeed, exist, will be shown below. Firstly, we consider the
asymptotic behaviour of such a transporting trajectory.  Given that
$q\rightarrow \infty$, the asymptotic energy of the washboard particle
is bounded by
\begin{equation}
 E_1(\infty)\ge E_1(0)-\frac{D}{\cosh(q_0-\widehat{Q})}\,.\label{eq:Easym}
\end{equation}
With the assumption in Eqs.\,(\ref{eq:E10}) and (\ref{eq:Ea}) we deduce that 
\begin{equation}
 E_1(\infty)\ge E_a>\frac{1}{\pi}\,.
\end{equation}
The last property means that the washboard particle loses only a
finite amount of energy and the energy of the washboard particle
approaches the level $E=E_a$ from above.  At the same time the energy
of the harmonic oscillator asymptotically approaches $E_2=E-E_a-D$
from below.  Hence, the coordinate of the harmonic oscillator performs
bounded oscillations around $Q=0$ which, together with growing $q$,
assures that the distance $q(t)-Q(t)$ effectively grows as time
progresses.  Hence, for initial coordinates fulfilling the constraints
of the theorem, the integrand in (\ref{eq:estimateexch}) vanishes in
an exponential fashion.

In conclusion, as soon as the inequalities
(\ref{eq:E10}),(\ref{eq:E20}) and (\ref{eq:qQ}), imposing conditions
in phase space, are satisfied, the process of redistributing the
finite (albeit small) amount of energy $D/\cosh(q-Q)$, that the
washboard degree of freedom possesses in excess of the asymptotic
level $E_a$, into the harmonic oscillator is {\it irreversible}. This
asymptotically decouples the two subsystems and, most importantly, the
washboard particle, being equipped with energy $E_1\ge E_a>1/\pi$,
moves in an unhindered and unidirectional manner, which establishes
continual transport that cannot be terminated.
\begin{figure}
\includegraphics[height=7cm, width=8cm]{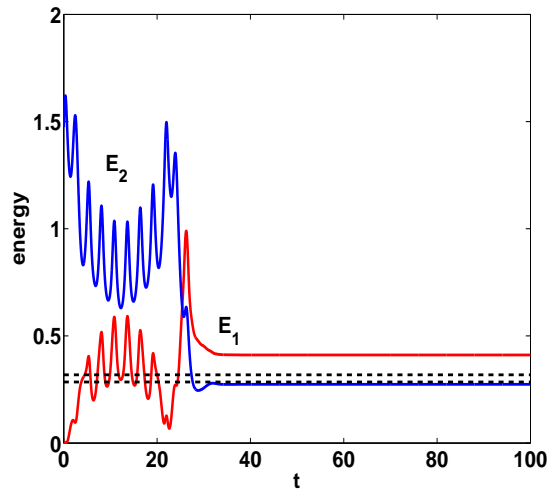}
\caption{Time evolution of the energy of the washboard particle $E_1$
  and the harmonic oscillator $E_2$.  The two dashed horizontal lines
  correspond to the energy levels $1/\pi$ and $E-D-E_a$ bounding the
  energies $E_1=E_a$ and $E_2=E-D-E_a$ in the asymptotic regime.} \label{fig:bottle}
\end{figure}
In Fig.~\ref{fig:bottle}, we illustrate a trajectory that shows an
initial energy redistribution from the harmonic oscillator into the
washboard particle, after which the phase space region conforming to
the assumptions of the theorem is reached.  Notice the pronounced
growth of the energy $E_1$ at the expense of $E_2$ for $t \gtrsim
25$. Crucially the energy $E_2$ falls below the level $ E-D-E_a$ at $t
\simeq 27$, that is at the {\it moment of no return}.  Strikingly,
this happens at the moment when the difference $q-Q$ starts to
increase beyond the value $\sinh^{-1}(1)$.  Afterwards the above
mentioned (small) energy loss (gain) in $E_1$ ($E_2$) takes place as a
result of which the energy $E_1$ settles onto $E_a$. The latter value
lies above the lower bound $1/\pi$ as given in (\ref{eq:Ea})
(represented by the lower dashed line in Fig.~\ref{fig:bottle}) while
the energy $E_2$ attains an asymptotic level below $E-D-E_a$
(represented by the upper dashed line in Fig.~\ref{fig:bottle}). 
%

In summary, we have demonstrated that it is possible to obtain
directed motion in autonomous Hamiltonian systems under the minimal
conditions that (i) transient chaos is supported and (ii) some of the
degrees of freedom serve to break the time reversibility with regard
to a set of unbiased localised initial conditions. (The model system
is minimal since these two conditions constitute the only
indispensable prerequisites. Notice in particular that no additional
modulation field is required in order to break the necessary
spatio-temporal symmetries.)  That chaos is needed only in an initial
stage of the dynamics in order to guide the trajectory into the range
of regular motion has a drastic implication, namely the directed net
motion is provided by regular motion and thus is of {\it continual}
nature.  This is at variance with previous studies of transport in
non-autonomous~\cite{Flach}--\cite{Schanz1} and autonomous Hamiltonian
systems~\cite{Acevedo},\cite{Schanz2}, where a mixing phase space is
needed to support directed chaotic transport. Despite the fact that
finite asymptotic currents were observed, sustaining transport in
general in nonintegrable systems seems to be hampered because of the
intermittent character of the dynamics inducing power-law decay
statistics of the duration of periods of regular motion.  On the other
hand, provided the minimal prerequisites mentioned above are given,
application of our approach to accomplish directed net motion in other
Hamiltonian systems, such as infinite lattice systems, is
straightforward. For example, in the context of the Holstein model,
the charge being initially trapped in a confined region of the
molecular chain constitutes localised initial conditions. Charge
motion along a molecular chain can be directed by taking into account
non-reversion symmetric oscillators representing the intra-molecular
vibrational degrees of freedom.

\end{document}